\documentclass[aps,prl,twocolumn,superscriptaddress,showpacs,floatfix]{revtex4-1}
  \usepackage{graphicx}
  \usepackage{epic,eepic}
  \usepackage{graphics}
  \usepackage{epsfig}
  \usepackage{subfigure}

  \newcommand{\be}{\begin{equation}}
  \newcommand{\ee}{\end{equation}}
  \newcommand{\bea}{\begin{eqnarray}}
  \newcommand{\eea}{\end{eqnarray}}

  \newcommand{\uv}{\mathbf{u}}

  \usepackage{amsmath}
  \usepackage{amsfonts}
  \usepackage{amssymb}
  \usepackage{bm}
  \usepackage{color}
  \usepackage{graphicx}

\begin{document}
  \title{Heat flux scaling in turbulent Rayleigh-B\'enard
    convection\\ with an imposed longitudinal wind}

\author{Andrea Scagliarini} \affiliation{School of Science and
  Engineering, Reykjavik University, Menntavegur 1, IS-101 Reykjavik,
  Iceland\\Department of Mathematics and Computer Science, Eindhoven
  University of Technology, The Netherlands\\Department of Physics and
  INFN, Univ. of Tor Vergata, Via della Ricerca Scientifica 1, 00133
  Rome, Italy}

\author{\'Armann Gylfason} \affiliation{School of Science and Engineering,
  Reykjavik University, Menntavegur 1, IS-101 Reykjavik, Iceland}

\author{Federico Toschi} 
\affiliation{Department of Mathematics and Computer Science, Eindhoven
  University of Technology, The Netherlands\\Department
  of Applied Physics, Eindhoven University of Technology, The
  Netherlands\\CNR-IAC, Via dei Taurini 19, 00185 Rome,
  Italy}

\begin{abstract}
We present a numerical study of Rayleigh-B\'enard convection disturbed
by a longitudinal wind. Our results show that under the action of the
wind, the vertical heat flux through the cell initially decreases, due
to the mechanism of plumes-sweeping, and then increases again when
turbulent forced convection dominates over the buoyancy. As a result,
the Nusselt number is a non-monotonic function of the shear Reynolds
number. We provide a simple model that captures with good accuracy all
the dynamical regimes observed. We expect that our findings can lead
the way to a more fundamental understanding of the of the complex
interplay between mean-wind and plumes ejection in the
Rayleigh-B\'enard phenomenology.
\end{abstract}

\date{}

\maketitle 
 
Thermal convection plays an important role in
many geophysical, environmental and industrial flows, such as in the
Earth's mantle, in the atmosphere, in the oceans, to name but a few
relevant examples.  In particular, the idealized case of
Rayleigh-B\'enard (RB) convection occurring in a layer of fluid
confined between two differentially heated parallel plates under a
constant gravitational field has been extensively studied
\cite{kadanoff,ahlersrev,lohsexia,schumachilla}.
However, in several real-life situations, the picture can be much more
complex with horizontal winds perturbing natural convection.  In the
atmosphere, for instance, this competition plays a crucial role in the
formation of thermoconvective storms \cite{bluestein}. On the other
side buoyancy effects can be relevant in a number of industrial
processes based on forced convection, such as coiled heat exchangers
\cite{steenhoven}. Similarly, a combination of forced and natural
convection is present in indoor ventilation applications
\cite{linden_review,bailon12,shishkina12}.

\begin{figure}[t]
\begin{center}
  \includegraphics[width=.73\hsize]{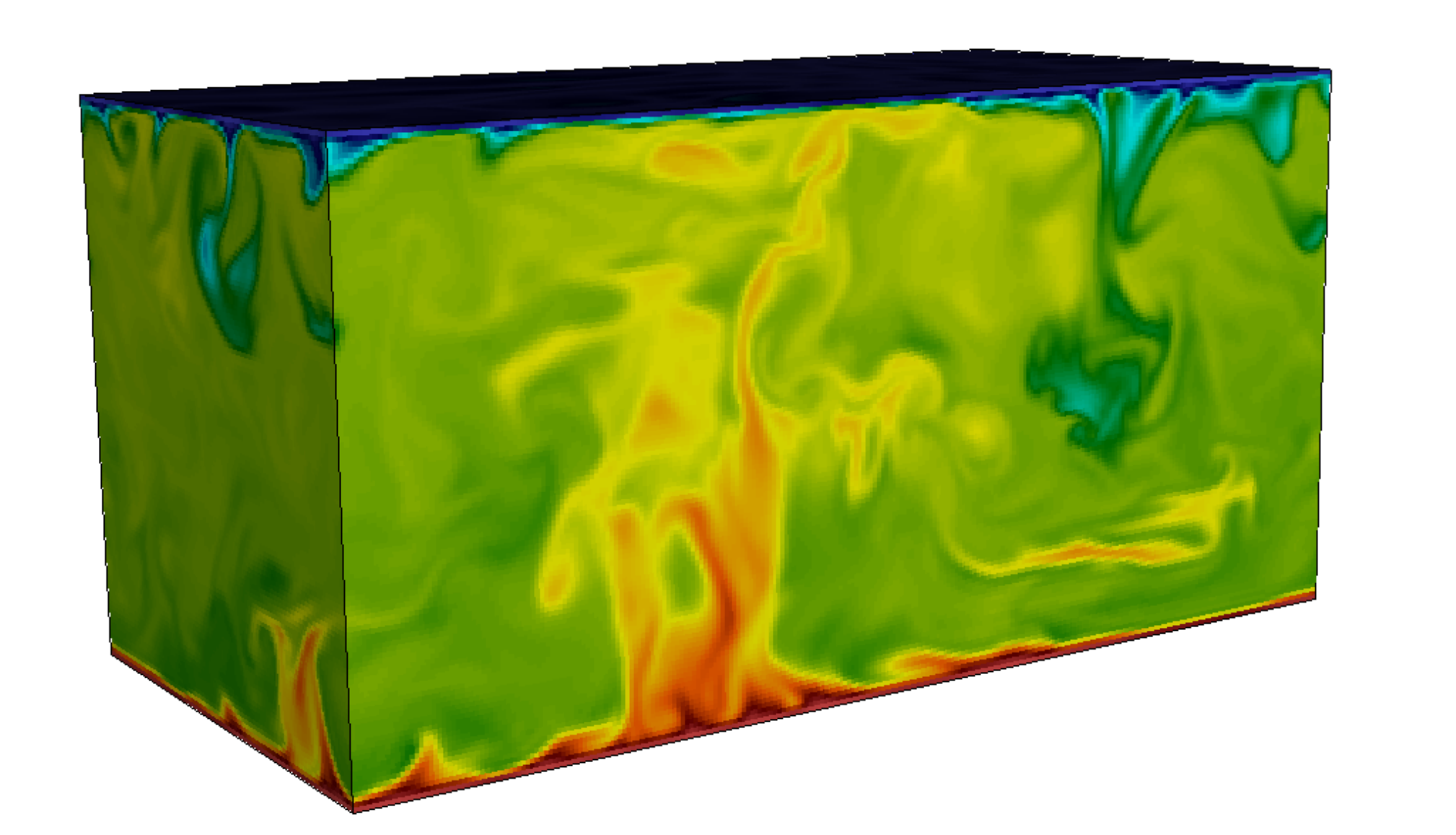}
 \label{fig0}
  \includegraphics[width=.73\hsize]{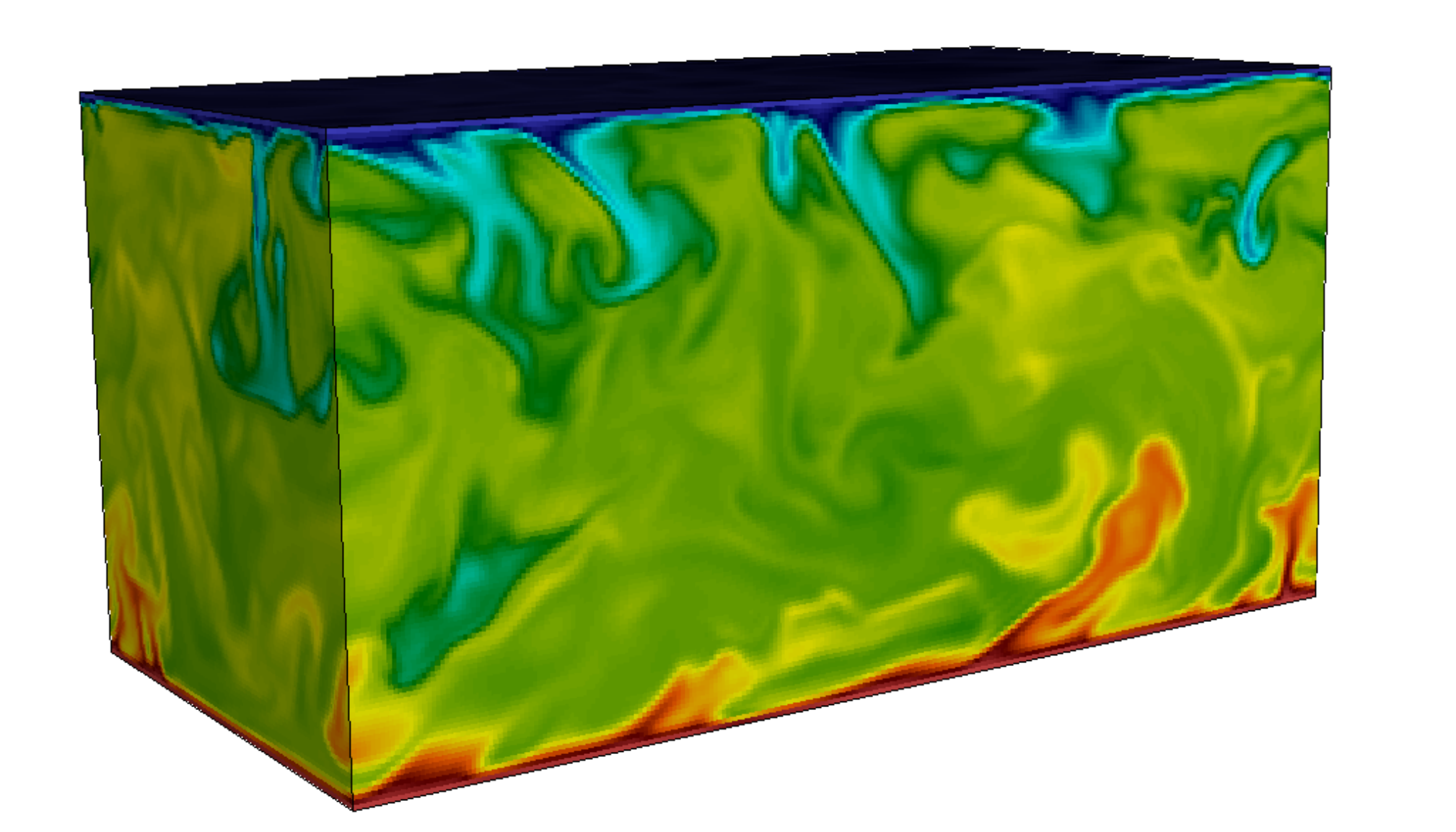}
 \caption{(top panel) Snapshot of the temperature field for the pure
   RB case ($Re_{\tau}=0$) at $Ra=1.3 \times 10^7$.  (bottom) Snapshot
   of the temperature field for the $Ra=1.3 \times 10^7$ and
   $Re_{\tau}=92$. Notice that, unlike figure \ref{fig0a}, where a
   buoyant plume detaching from the bottom boundary layer can easily
   enter the bulk up to the top plate while, here plumes are
   considerably distorted in the direction of the wind.}
\end{center}
\end{figure}
According to the standard picture at the basis of
the existing models for the scaling laws of the heat flux
\cite{shraiman,GL00,GL04}, the RB system is characterized by the
multi-scale coupling between large-scale circulation (mean wind) and
detaching thermal structures from the boundary layers at the walls
(plumes). Besides the above mentioned motivations, applying a mean
wind to a natural convection setup can shed light on the effectes of
bulk flow on the boundary layer dynamics, thus helping to better
understand one of the most intriguing feature of RB convection.\\ In
this Letter we report on a numerical study of RB convection with an
imposed constant horizontal pressure gradient, orthogonal to gravity,
that induces the wind (the so-called Poiseuille-Rayleigh-B\'enard
(PRB) flow setup \cite{gage,carriere}).  We show that the heat
transfer from the walls can be dominated by either the buoyancy or by
the ``forced'' convection and that the interplay of the two mechanisms
gives rise to  a non-trivial dependence of the Nusselt number, $Nu$,
on the parameter space that is spanned by the Rayleigh, $Ra$, and
shear Reynolds numbers, $Re_{\tau}$ (quantifying, respectively, the
intensity of buoyancy and of the pressure gradient relatively to
viscous forces).\\ Our main result consists in the observation that,
taken a standard RB system as reference, $Nu$ initially decreases and
then, when the dynamics is completely dominated by the forced
convection regime, it increases again with $Re_{\tau}$.  A
phenomenological explanation for this behaviour is provided together
with discussions on the possible implications for the modelling of the
$Nu \; vs \; Ra$ relation in pure natural convection setup.

The equations of motion for the fluid velocity, $\uv$, and temperature, $T$,
 are:
\begin{equation} \label{eq:NS} 
\partial_t \uv + \mathbf{u}\cdot \nabla\uv= -\frac{1}{\rho}\nabla P +
\nu \nabla^2 \uv + \alpha\mathbf{g}T + \mathbf{f}
\end{equation} 
\begin{equation} \label{eq:temp} 
\partial_t T + \mathbf{u}\cdot \nabla T = \kappa \nabla^2 T,
\end{equation}
in addition to the incompressibility condition, $\nabla \cdot
\uv$. The properties of the fluid are $\rho$ the (assumed constant)
fluid density, $\nu$ the kinematic viscosity, $\alpha$ the thermal
expansion coefficient, and $\kappa$ the thermal diffusivity. $P$ is
the pressure field, $\mathbf{g}=g \hat{z}$ the gravity and
$\mathbf{f}$ a forcing term of the form $\mathbf{f} = (F/\rho) \hat{x}
\equiv \tilde{F} \hat{x}$ ($\hat{x}$ is the direction parallel to the
walls, or stream-wise direction).  Equations (\ref{eq:NS}) and
(\ref{eq:temp}) are evolved using a 3d thermal lattice Boltzmann
algorithm \cite{succi,wolf-gladrow} with two probability densities
(for density/momentum and for temperature, respectively).  As
mentioned in the introduction, to characterize the dynamics we need
two parameters: the Rayleigh number, $Ra$, quantifying the strength of
buoyancy (with respect to viscous forces),
$$
Ra= \frac{\alpha g \Delta H^3}{\nu \kappa},
$$ (where $\Delta = T_{hot} - T_{cold}$ is the temperature drop across
the cell and $H$ the cell height), and the shear Reynolds number
$Re_{\tau}$,
$$Re_{\tau} = {H \over {2\nu}}\sqrt{{\tilde{F} H} \over 2}.$$ We
performed several runs (in a computational box of size $256 \times 128
\times 128$, uniform grid; see figure \ref{fig0}  for
snapshots of the temperature field in the simulation cell)), exploring
the two dimensional parameter space $(Ra, Re_{\tau})$, within the
ranges $Ra \in [0; 1.3 \times 10^7]$ and $Re_{\tau} \in [0; 205]$; the
Prandtl number $Pr=\nu/\kappa$ is kept fixed and equal to one.

\begin{figure}[!t]
\begin{center}
  \advance\leftskip-0.55cm
  \includegraphics[scale=0.40]{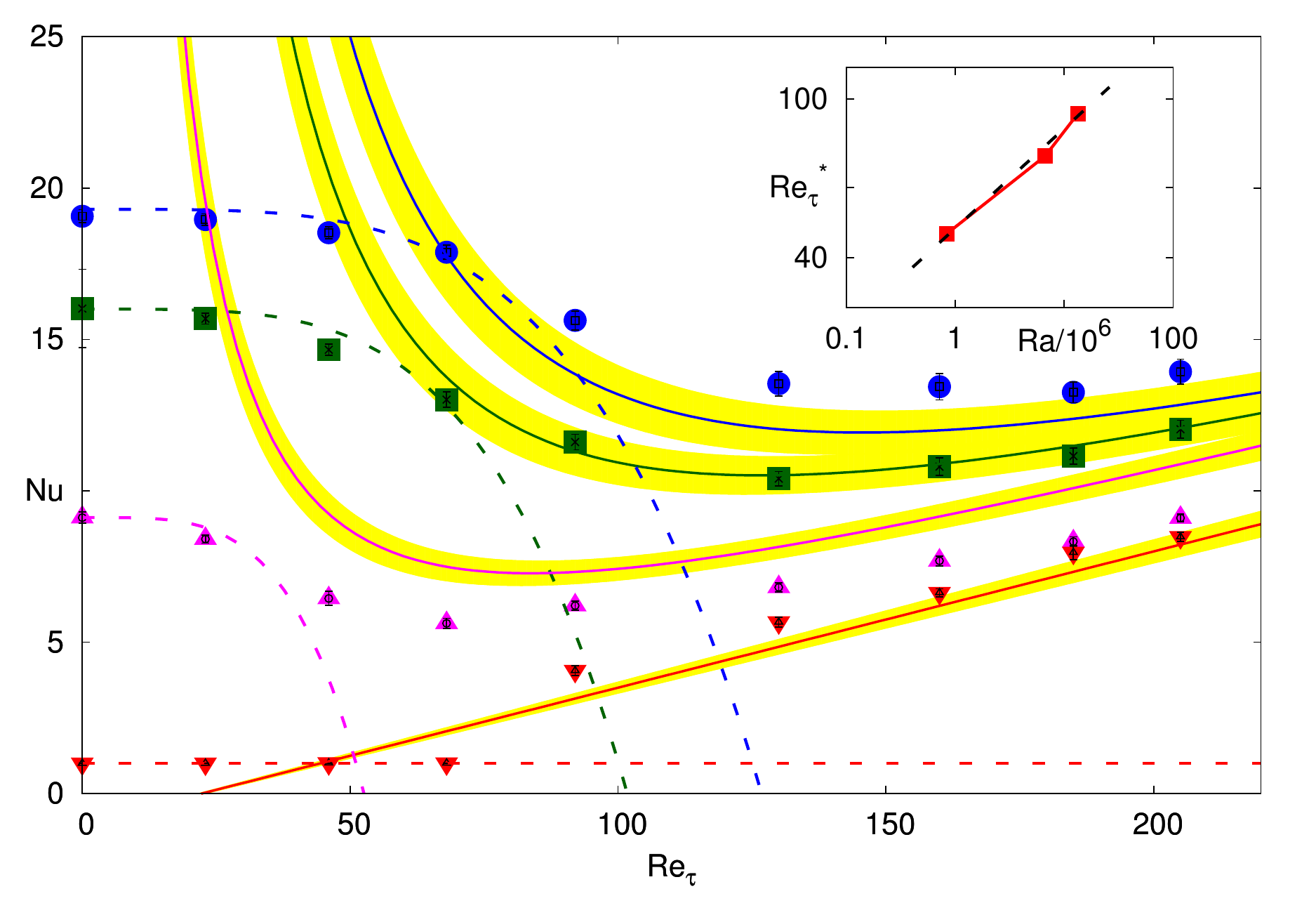}
  \caption{Nusselt number as a function of $Re_{\tau}$ for $Ra=0,
    8.125 \times 10^5, 6.5 \times 10^6, 1.3 \times 10^7$ (from bottom
    to top, symbols respectively $\blacktriangledown$,
    $\blacktriangle$, $\blacksquare$, $\bullet$). Notice the
    non-monotonic behaviour for $Ra>0$: for low-to-moderate
    $Re_{\tau}$ a decrease of $Nu$ is observed due to plumes-sweeping
    by the mean wind; at higher $Re_{\tau}$, when bouyancy is
    basically irrelevant with respect to the channel flow, $Nu$
    increases again, as expected for turbulent forced convection. The
    solid lines are the predictions from equation (\ref{eq:model})
    obtained using $A_1=2.1$ and $A_2=0.045$, values that have been
    chosen to best fit the case $Ra =6.5 \times 10^6$ ($\blacksquare$
    symbols). The yellow band around the solid lines show the
    prediction of the model for a variation of the parameters $A_1$
    and $A_2$ of $\pm 10\%$.
The dashed lines represent the fall-off for small $Re_{\tau}$ provided
by equation (\ref{eq:Ret4}), with $A_3 = 1$ for all the three $Ra$.
The reader is referred to the text for the details. (Inset) Shear
Reynolds number $Re_{\tau}^{\ast}$ corresponding to the crossover to
the $Nu \sim Re_{\tau}^{-3/2}$ regime as function of Rayleigh
number. The dashed line is the $Re_{\tau}^{\ast} \sim Ra^{1/4}$ power
law, equation (\ref{eq:crossover}).}
\label{fig1}
\end{center}
\end{figure}

\begin{figure}[!t]
\begin{center}
  \advance\leftskip-0.55cm
  \includegraphics[scale=0.7]{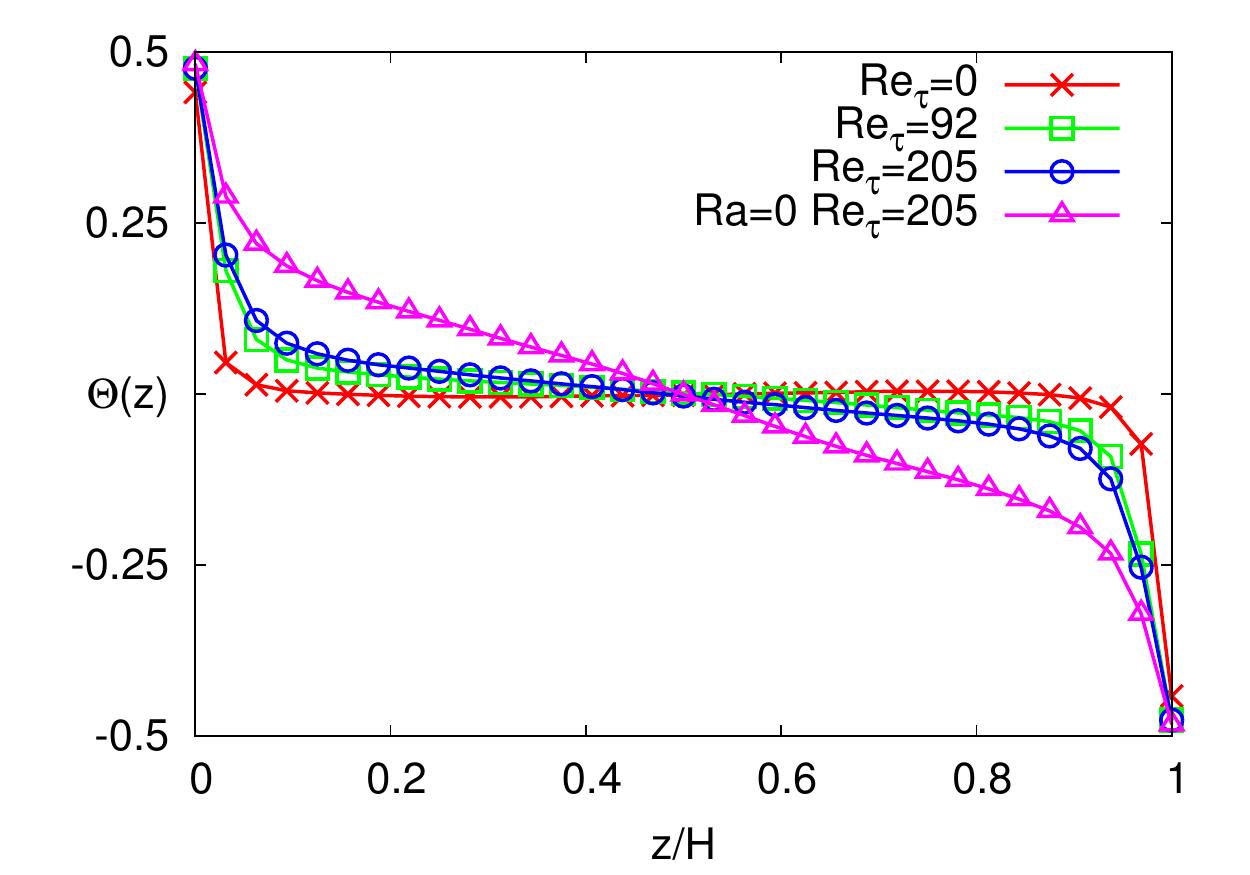}
  \caption{Normalized temperature profiles $\Theta(z) \equiv
    \frac{\overline{T}(z) - T_m}{\Delta}$ ($T_m=(T_{hot}+T_{cold})/2$
    being the mean temperature) for various $Re_{\tau}$ at a fixed
    $Ra=6.5 \times 10^6$. Notice the bending of the profile in the
    bulk in comparison with the usual 'thermal short-cut' for
    $Re_{\tau} = 0$ (i.e., no lateral wind).}
\label{fig2}
\end{center}
\end{figure}

\begin{figure}[!t]
\begin{center}
  \advance\leftskip-0.55cm
 \includegraphics[scale=0.7]{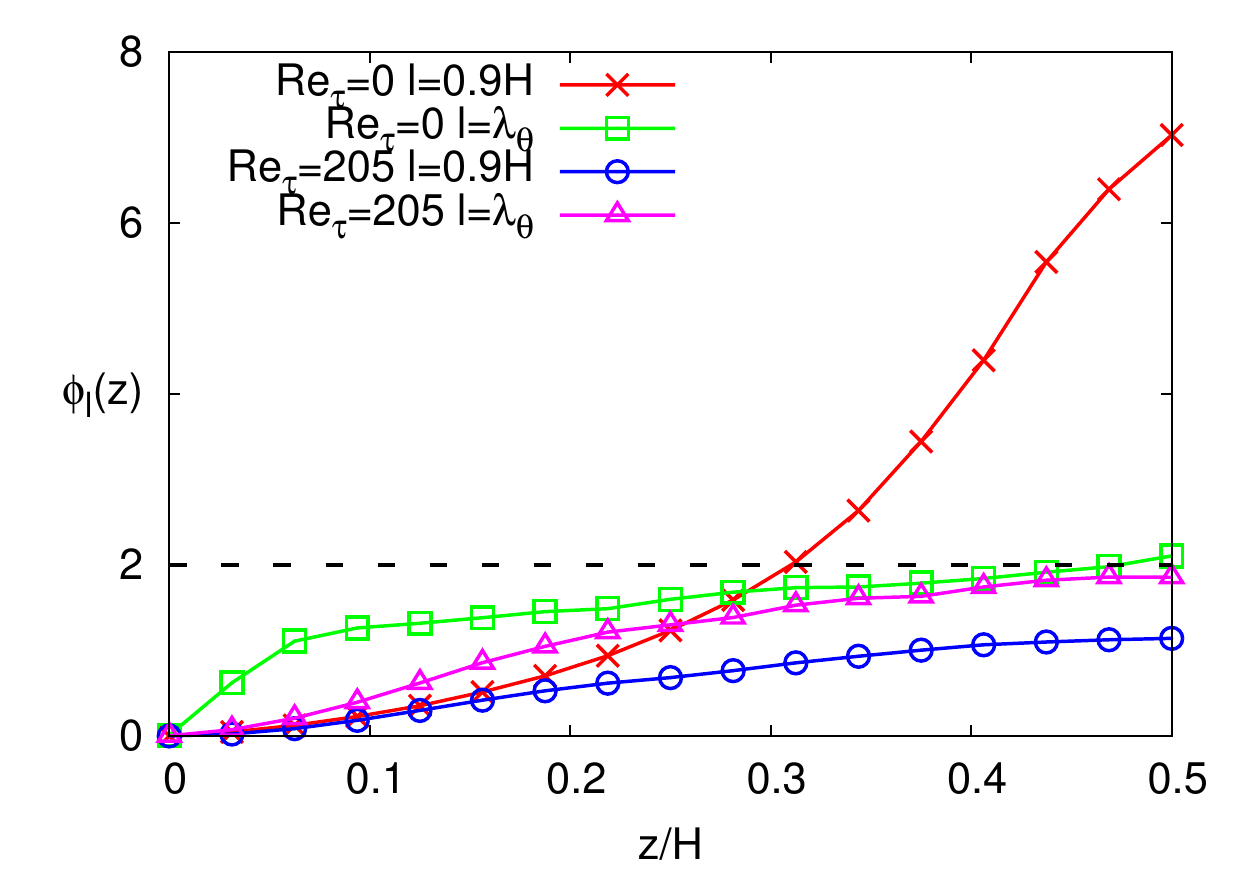}
  \caption{The anisotropy indicator defined in Eqn. (\ref{eq:aniso})
    as function of the cell height for pure RB ($Re_{\tau}=0$) and for
    $Re_{\tau}=205$ and for two separation in $x$, at $Ra = 6.5 \times
    10^6$.}
\label{fig3}
\end{center}
\end{figure}

A typical key question in RB studies is how the dimensionless heat
flux through the cell, $Nu$, varies as a function of the Rayleigh
number:
\begin{equation} \label{eq:nudef}
Nu(z) = \frac{\overline{u_z T}(z) - \kappa \partial_z
  \overline{T}(z)}{\kappa \frac{\Delta}{H}} = const \equiv Nu
\end{equation}
with $Ra$; here and hereafter the overline indicates a spatial (over
planes $z=const$) and temporal (over the statistically stationary
state) average.  The second and third equalities (which state that
$Nu$ is constant with $z$) follow from taking the average of equation
(\ref{eq:temp}).  In our setup in addition to buoyancy there is the
longitudinal pressure gradient which affects the heat flux.  We
therefore focus on the dependence of $Nu$ on the two-dimensional
parameter space $(Ra, Re_{\tau})$; in figure \ref{fig1} we plot $Nu$
as a function of $Re_{\tau}$ for various fixed $Ra$.  We find that,
for moderate/high $Ra$, the effect of the lateral wind is to quench
the buoyancy driven convection, and thus $Nu$ decreases with
$Re_{\tau}$. For very low $Ra$, below the critical Rayleigh number
$Ra_c$, the dynamics of the flow is instead completely dominated by
the forced convection and thus $Nu$ increases with the $Re_{\tau}$ (In
Fig. \ref{fig1} we show the $Ra=0$ case).  Correspondingly, at
increasing $Re_{\tau}$ the mean temperature profiles (see figure
\ref{fig2}) show a bending in the bulk and a decrease of the gradient
in the boundary layer.
Our interpretation of these observations is that, for small
$Re_{\tau}$/high $Ra$ (i.e. in the natural convection dominated
regime), the wind acts essentially sweeping away thermal plumes (which
are mixed and lose their coherence closer to the walls) and hence the
heat flux is depleted.  Increasing $Re_{\tau}$ more and more we
eventually reach a state where buoyancy becomes irrelevant. Here, $Nu$
starts to increase again by resuspension of temperature puffs in the
bulk due to bursts from the wall emerging because of the turbulent
channel flow.  To give an indication of the validity of such a
conjecture we have measured the following quantity:
\begin{equation} \label{eq:aniso}
\phi_{\ell}(z) \stackrel{def}{=} \frac{\overline{(\delta_{\ell}
    u_z)^2}}{\overline{(\delta_{\ell} u_x)^2}},
\end{equation}
where $\delta_{\ell} u_i \equiv u_i(x+\ell,y,z;t)- u_i(x,y,z;t)$.  The
observable (\ref{eq:aniso}) is the ratio of a generalized second order
transverse over longitudinal structure function and, as such, it
serves as a sort of scale-dependent anisotropy indicator: a large
value of $\phi_{\ell}(z)$ means a coherent motion in the wall-normal
direction.  In figure \ref{fig3} we plot $\phi_{\ell}(z)$ on a large
scale ($\ell \approx H$) and on a scale of the order of the thermal
boundary layer thickness ($\ell = \lambda_{\theta}$), which gives an
estimate of a characteristic size of plumes, for natural convection
($Ra=6.5 \times 10^6$, $Re_{\tau}=0$) and for a case with the wind
($Ra=6.5 \times 10^6$, $Re_{\tau}=205$). For the pure RB case
$\phi_{\ell \approx H}(z)$ grows to large values in the bulk, due to
the thermal wind, while $\phi_{\ell = \lambda_{\theta}}(z)$ goes to
the isotropic value $\phi \approx 2$ in the bulk and it is larger than
$\phi_{\ell \approx H }(z)$ close to the wall, pointing out the
presence of detaching plumes. The same quantity $\phi_{\ell =
  \lambda_{\theta}}(z)$ in the wall-proximal region is significantly
smaller for $Re_{\tau}=205$, indicating the depletion of plumes
ejection.

With this picture in mind we are now going to build a model to recover
the numerical findings. Our argument goes as follows. As shown in
figure \ref{fig2}, under the action of the lateral wind the
temperature profile ceases to be flat in the bulk.  This permits us to
write a first order closure for the turbulent heat flux of the kind:
\begin{equation} \label{eq:closure}
\overline{u_z T} = - \kappa_T \partial_z \overline{T}, 
\end{equation}
 where $\kappa_T$ is a turbulent diffusivity. When writing
(\ref{eq:closure}), where $\kappa_T$ is constant with $z$, we are
implicitly restricting ourselves to the bulk region (where the mean
temperature gradient is basically constant); we are allowed to do that
by (\ref{eq:nudef}), i.e. the constancy of the heat flux through
planes parallel to the walls.  The Nusselt number will assume the form
\begin{equation} \label{eq:nusselt}
Nu \sim \left(1 + \frac{\kappa_T}{\kappa} \right)\frac{\left| \partial_z \overline{T} \right|}{(\Delta/H)}.
\end{equation}
We consider that two types of structures contribute to turbulent
diffusion, namely buoyant plumes ($\kappa_T^{(P)}$) and bursts ($\kappa_T^{(B)}$, triggered by the
turbulent channel flow), so that we may write 
\begin{equation}
\kappa_T = \kappa_T^{(P)}+ \kappa_T^{(B)}. 
\end{equation}
As previously discussed we attribute the heat flux reduction to the
sweeping of plumes by the wind; we model this saying that the plume
looses its coherence (or else, it releases its heat content) after
travelling a distance from the wall of the order of the kinetic
boundary layer thickness ($\lambda_u$), i.e. we suggest that we can
adopt a Prandtl mixing length ($\ell_m$) theory kind of approach,
using
\begin{equation}
\ell_m  \sim \lambda_u; 
\end{equation}
the latter relation should be interpreted as a scaling (or
proportionality) relation rather than an order of magnitude.  The
characteristic velocity of a rising plume reaching a height $\sim
\ell_m$ can be estimated as $u \sim \sqrt{\alpha g \Delta \ell_m}$
\cite{footnote1}, hence the contribution to the turbulent diffusion
will be
\begin{equation} 
\kappa_T^{(P)} = \sqrt{\alpha g \Delta} \lambda_u^{3/2}
\end{equation} 
and assuming a laminar boundary layer of Blasius type 
of thickness \cite{landau}
\begin{equation} \label{eq:blasius}
\lambda_u \sim \frac{H}{Re_{\tau}}
\end{equation}
we get
\begin{equation} \label{eq:Kplumes}
\kappa_T^{(P)}\sim \sqrt{\alpha g \Delta} \frac{H^{3/2}}{Re_{\tau}^{3/2}}.
\end{equation}
For a turbulent burst one can also assume that $\ell_m \sim
\lambda_u$, but the expression for the characteristic advecting
velocity requires some more care. In a pure forced convection setup
(or in our case when the wind is dominant) there is no buoyancy, so we
cannot use the expression of the free-fall velocity; instead
convection is driven by turbulent fluctuations from the wall. Invoking
again the mixing length theory for a first order closure for the
velocity we can write $u_z \sim \ell_m \partial_z \overline{U}_x$,
whence 
\begin{equation}
\kappa_T^{(B)} \sim \ell_m^2 \partial_z \overline{U}_x \sim \lambda_u^2  \partial_z \overline{U}_x;
\end{equation} 
estimating the shear as $\partial_z U_x \sim U_c/\lambda_u$ ($U_c$
being the centreline velocity) we get
\begin{equation} 
\kappa_T^{(B)}\sim \lambda_u U_c.
\end{equation} 
In the limited range of $Re_{\tau}$ that we span it is reasonable to
assume that the friction coefficient goes as $C_f \sim
Re_{\tau}^{-2}$, hence that $U_c$ scale as $U_c \sim (\nu / H )Re_{\tau}^2$
\cite{footnote2}.  Inserting this scaling law together with the
relation (\ref{eq:blasius}) inside the expression for $\kappa_T^{(B)}
$ we obtain
\begin{equation} \label{eq:Kbursts}
\kappa_T^{(B)} \sim \nu \cdot  Re_{\tau}.
\end{equation} 
Putting the expressions (\ref{eq:Kplumes}) and (\ref{eq:Kbursts})
inside equation (\ref{eq:nusselt}) we end up with
\begin{equation}
Nu -1 \sim \left( A_1\frac{\sqrt{\alpha g \Delta}
    H^{3/2}}{\kappa Re_{\tau}^{3/2}} + A_2 \frac{\nu}{\kappa}  Re_{\tau} \right)
\frac{\left| \partial_z \overline{T} \right|}{(\Delta/H)},
\end{equation}
which can be recast, introducing the dimensionless numbers $Ra$ and
$Pr$, into the following form:
\begin{equation} \label{eq:model}
Nu -1\sim \left( A_1\frac{Ra^{1/2}Pr^{1/2}}{Re_{\tau}^{3/2}} + A_2 Pr
  Re_{\tau} \right)
\frac{\left| \partial_z \overline{T} \right|}{(\Delta/H)},
\end{equation} 
where $A_1$ and $A_2$ are two free parameters of the model.  Some
comments on equation (\ref{eq:model}) are in order. Firstly, it
reproduces the non-monotonic dependence of the heat flux, $Nu$, on the
applied wind, $Re_{\tau}$, and it turns out to be in fair agreement
with the numerical data (see figure \ref{fig1}).  Secondly, it
provides an argument for the scaling $Nu \sim Re_{\tau}$ for the case
of pure forced convection ($Ra=0$, see figure \ref{fig1}).  It is
interesting to note that, for very small $Re_{\tau}$, our model would
give a scaling $Nu \sim Ra^{1/2}$, i.e. what expected for Kraichnan's
{\it ultimate regime} of convection \cite{ahlersrev}.

The phenomenology behind it suggests that the lower heat flux
observed in the standard RB convection (with respect to $Ra^{1/2}$)
may be seen as the result of a negative feedback of the shear, due to
the large scale circulation, on the plumes detaching from the boundary
layer.  Indeed, if we imagine the Nusselt number to follow a Kraichnan
scaling on an {\it effective} Rayleigh $Ra_{eff}$, renormalized by
turbulent viscosity and thermal diffusivity (behaving as
(\ref{eq:Kbursts})), that is
$$
Ra_{eff} = \frac{Ra}{(\nu_T/\nu)(\kappa_T/\kappa)},
$$
we end up with the following relation
\begin{equation} \label{eq:effsca}
Nu \sim Ra_{eff}^{1/2} \equiv \frac{Ra^{1/2}}{Re_{\tau}}.
\end{equation}
If we now insert into (\ref{eq:effsca}) the ultimate regime scaling
for Reynolds $Re_{\tau} \sim Ra^{1/4}$ \cite{footnote3}, we obtain
\begin{equation}
Nu \sim Ra^{1/4},
\end{equation}
a well known scaling, predicted theoretically and found in a vast
number of experiments (see \cite{GL00} and references therein).  Let
us, finally, remark that equation (\ref{eq:model}) should not be
expected to be valid for $Re_{\tau} \rightarrow 0$, since in this case
the mean temperature gradient is zero and a closure like
(\ref{eq:closure}) does not apply \cite{footnote4}.  In particular we
detect a region where the sweeping mechanism is not yet effective and
$Nu$ decreases slowly with $Re_{\tau}$; we denote the shear Reynolds
number at which the crossover between such region and the $Nu \sim
Re_{\tau}^{-3/2}$ regime takes place as $Re_{\tau}^{\ast}$ and we
argue that such crossover can be determined under the condition that
the characteristic velocity of a rising plume, $U^{(RB)} \sim
\sqrt{\alpha g \Delta H}$, be of the same order of the centreline
velocity of the Poiseuille flow, $U^{(P)} \sim \tilde{F}
H^2/\nu$. Equating these two latter relations we have
$$
\sqrt{\alpha g \Delta H} \sim \frac{ \tilde{F} H^2}{\nu},
$$
which gives, in dimensionless form and introducing the crossover Reynolds,
\begin{equation} \label{eq:crossover}
Re_{\tau}^{\ast} \sim Ra^{1/4}.
\end{equation}
This results is compared with the numerical data in the inset of
figure \ref{fig1}.
For $Re_{\tau} < Re_{\tau}^{\ast}$ the longitudinal flow is still laminar (notice that
the Nusselt number for $Ra=0$ remains equal to one) and represents
just a small disturbance to the buoyant circulation. The initial
fall-off of $Nu$ vs $Re_{\tau}$ can be captured by looking at the
conservation equation for the total energy, which can be derived
from (\ref{eq:NS}) and (\ref{eq:temp}) to be \cite{shraiman}
\begin{equation} \label{eq:enecons}
\varepsilon = (Nu-1)Ra + 8 Re_{\tau}^2 \langle u_x \rangle,
\end{equation}
where $ \langle \cdots \rangle$ denotes an average over the entire
volume, $\varepsilon = \langle (\partial_i u_j)^2 \rangle$ is the
kinetic energy dissipation rate and we set $Pr=1$.  It is clear that
$\langle u_x \rangle \sim U^{(P)} \sim Re_{\tau}^2$ ($\langle U^{(RB)}
\rangle \sim 0$). Since $U^{(RB)} \gg U^{(P)}$ the longitudinal wind
perturbs the RB dynamics only slightly so that $\varepsilon \approx
\varepsilon^{(RB)}$ \cite{footnote5}. From (\ref{eq:enecons}) we
therefore derive
\begin{equation} \label{eq:Ret4}
Nu \approx Nu_0(Ra) - (A_3/Ra) Re_{\tau}^4,
\end{equation}
where $Nu_0$ is the Nusselt number for $Re_{\tau}=0$, i.e. pure RB,
and $A_3$ is an order one constant. Equation (\ref{eq:Ret4}) is
plotted in figure \ref{fig1} for three different $Ra$ showing good
agreement with the numerics up to the expected crossover shear
Reynolds $Re_{\tau}^{\ast}$.

We have performed direct numerical simulation of Rayleigh-B\'enard
convection with an imposed longitudinal pressure gradient inducing a
mean wind. We found that the Nusselt number has a non-monotonic
dependence on the shear Reynolds number based on the applied pressure
drop: to an initial decrease (justifiable in terms of a mechanism of
sweeping of plumes by the longitudinal wind) an increase follows, when
the dynamics is dominated by the turbulent ``forced convection''
regime. Based on these empirical concepts, we provided a correlation
which proved able to recover the numerical findings with reasonable
accuracy. The observations and the modelling give a hint that, in
standard RB convection, the shear due to the large scale circulation
may act back onto the boundary layer against the ejection of plumes to
the bulk (thus being a possible mechanism for the depletion of heat
transfer respect to the ultimate state of turbulent convection).  Our
work is a first attempt to look directly at the effect of disturbing
in a controlled manner the dynamics of the boundary layer in such a
way to give an insight of its role in natural convection. A possible
follow-up of the present study is to use a perturbation other than a
simple Poiseuille flow.

{\it Acknowledgements}.  We thank R. Benzi, P. Roche, R.P.J. Kunnen, F. Zonta and
P. Ripesi for useful discussions and L. Bouhlali for careful reading
of the manuscript. AS and AG acknowledge financial support from the
Icelandic Research Fund. AS acknowledges FT and the Department of
Mathematics and Computer Science of the Eindhoven University of
Technology for the hospitality.


\addcontentsline{toc}{section}{References}

\begin{thebibliography}{99}

\bibitem{kadanoff} L.P. Kadanoff, ``Turbulent Heat Flow: Structures
  and Scaling'', {\it Physics Today} 34-39 (2001) .

\bibitem{ahlersrev} G. Ahlers, S. Grossmann and D. Lohse, {\it
    Rev. Mod. Phys.} {\bf 81}, 503 (2009).

\bibitem{lohsexia} D. Lohse and K.-Q. Xia, 
{\it  Annu. Rev. Fluid Mech.} {\bf 42}, 335 (2010).

\bibitem{schumachilla} J. Schumacher and F. Chill\`a, {\it
  Eur. Phys. J. E} {\bf 35}, 58 (2012).

\bibitem{bluestein} H.B. Bluestein,
{\it Severe convective storms and tornadoes},
Springer (2013).

\bibitem{steenhoven} J.J.M. Sillekens, C.C.M. Rindt and A.A. Van Steenhoven,
{\it Int. J. Heat and Mass Transf.} {\bf 41}, 61 (1998).

\bibitem{linden_review} P.F. Linden, {\it Annu Rev Fluid Mech} {\bf
  31} 201 (1999).

\bibitem{bailon12} J. Bailon-Cuba, O. Shishkina, C. Wagner, and
  J. Schumacher {\it Phys Fluids} {\bf 24} 107101 (2012).

\bibitem{shishkina12} O. Shishkina and C. Wagner, {\it J Turbul} {\bf
  13} N22 (2012).
 
\bibitem{shraiman} B.I. Shraiman and E.D. Siggia, {\it Phys. Rev. A}
  {\bf 42}, 3650 (1990).

\bibitem{GL00} S. Grossmann and D. Lohse, {\it J. Fluid Mech.}
  {\bf 407}, 27 (2000).

\bibitem{GL04} S. Grossmann and D. Lohse, {\it Phys. Fluids}
  {\bf 16}, 4462 (2004).

\bibitem{gage} K.S. Gage and W.H. Reid, {\it J. Fluid Mech.}  
{\bf 33}, 21 (1968).

\bibitem{carriere} P. Carri\`ere, P.A. Monkewitz and D. Martinand, {\it J. Fluid Mech.}  
{\bf 502}, 153 (2004).

\bibitem{succi} S. Succi, {\it The Lattice Boltzmann equation for
    Fluid Dynamics and beyond}. Oxford University Press (2001).

\bibitem{wolf-gladrow} D. Wolf-Gladrow, {\it Lattice Gas Cellular
    Automata and Lattice Boltzmann Methods}. Springer (2000).

\bibitem{footnote1} We approximate that the plume still feels an
  acceleration proportional to the temperature difference $\Delta$,
  i.e. that the bending of the thermal short-cut is small.

\bibitem{landau} L.D. Landau and E.M Lifshitz, {\it Fluid
    Mechanics}. Pergamon Press (1959).

\bibitem{footnote2} More precisely our simulations suggest something
  closer to $U_c \sim Re_{\tau}^{1.9}$.

\bibitem{footnote3} The scaling is $Re \sim Ra^{1/2}$ and then, since
  $Re_{\tau} \sim Re^{1/2}$, we get $Re_{\tau} \sim Ra^{1/4}$.

\bibitem{footnote4} In principle there would be an extra dependence of
  $Nu$ on $Re_{\tau}$ in equation (\ref{eq:model}) stemming from
  $\left|\partial_z \overline{T} \right|/(\Delta/H) \sim
  f(Re_{\tau})$, which, however, turns out to be a subdominant
  correction, only becoming relevant for very low $Re_{\tau}$.

\bibitem{footnote5} We assume that the energy dissipation rate is not
  affected by the longitudinal wind, at least in its boundary layer
  contributions which are dominant in this regime (our numerical simulations confirm this picture).



\end{thebibliography}
\end{document}